\documentclass[pdflatex,sn-mathphys-num]{sn-jnl}

\usepackage{graphicx}%
\usepackage{multirow}%
\usepackage{amsmath,amssymb,amsfonts}%
\usepackage{amsthm}%
\usepackage{mathrsfs}%
\usepackage[title]{appendix}%
\usepackage[dvipsnames]{xcolor}%
\usepackage{xspace}
\usepackage{textcomp}%
\usepackage{manyfoot}%
\usepackage{booktabs}%
\usepackage{algorithm}%
\usepackage{algorithmicx}%
\usepackage{algpseudocode}%
\usepackage{listings}%
\usepackage{makecell}  
\usepackage{caption}
\usepackage{subcaption}
\usepackage{graphicx}
\usepackage[newenum,newitem]{paralist}
\usepackage[capitalize]{cleveref}
\usepackage{array}

\newcommand{\NP}{Network Pluralism\xspace}
\theoremstyle{thmstyleone}%

\theoremstyle{thmstyletwo}%

\theoremstyle{thmstylethree}%

\begin{document}

\title[The Power of Network Pluralism]{The Power of Network Pluralism: Multi-Perspective Modeling of Heterogeneous Legal Document Networks}

\author*[1]{\fnm{Titus} \sur{P\"under}}\email{titus.puender@aalto.fi}

\author*[1,2,3]{\fnm{Corinna} \sur{Coupette}}\email{corinna.coupette@aalto.fi}

\affil[1]{\orgdiv{Department of Computer Science}, \orgname{Aalto University}, \orgaddress{\country{Finland}}}

\affil[2]{\orgdiv{Max Planck Institute for Tax Law and Public Finance}, \orgaddress{\country{Germany}}}

\affil[3]{\orgdiv{CodeX -- The Stanford Center for Legal Informatics}, \orgname{Stanford University}, \orgaddress{\country{United States}}}

\abstract{

Insights are relative---%
influenced by a range of factors such as assumptions, scopes, or methods that together define a \emph{research perspective}.
In normative and empirical fields alike,
this insight has led to the conclusion that no single perspective can generate complete knowledge.
As a response, \emph{epistemological pluralism} mandates that researchers consider multiple perspectives simultaneously to obtain a holistic understanding of their phenomenon under study.
Translating this mandate to network science, our work introduces
\emph{Network Pluralism} as a conceptual framework that leverages \emph{multi-perspectivity} to yield more complete, meaningful, and robust results.
We develop and demonstrate the benefits of this approach via a hands-on analysis of \emph{complex legal systems},
constructing a \emph{network space} from references across documents from different branches of government
as well as including organizational hierarchy above and fine-grained structure below the document level.
Leveraging the resulting heterogeneity in a \emph{multi-network analysis},
we show how \emph{complementing} perspectives can help contextualize otherwise high-level findings,
how \emph{contrasting} several networks derived from the same data enables researchers to \emph{learn by difference},
and how relating metrics to perspectives may increase the transparency and robustness of network-analytical results. 
To analyze a space of \emph{networks as perspectives}, 
researchers need to map \emph{dimensions of variation} in a given domain to network-modeling decisions and network-metric parameters. 
While this remains a challenging and inherently interdisciplinary task, 
our work acts as a blueprint to facilitate the broader adoption of Network Pluralism in domain-driven network research. 

}

\keywords{Legal Network Science, Epistemological Pluralism, Complex Systems}

\maketitle

\section{Introduction}

\emph{It depends}.
Two words frequently uttered by lawyers when confronted with a legal question.
\emph{It's complicated}.
Phrases that fail to satisfy the inquirer's desire for immediate certainty
and that are common starting points for caricatures and stereotypes of the legal profession.
\emph{Why have codified rules if they cannot give a definite answer?}, one might rightfully ask.
In search of certainty,
one might turn to the world of numbers and ask computer scientists, physicists, or (applied) mathematicians. 
Soon, however,
one will notice that their answers also depend on something, e.g., on prior assumptions.
There are only few situations where \emph{it does not depend} on something!
 
The general problem that \emph{knowledge depends} is by no means new. 
Rather, it has been addressed by thinkers and researchers in many fields and professions.
Some strive for a source of \emph{absolute} knowledge or some methodology to reach it.
Others have acknowledged the \emph{relativity} of knowledge to some \emph{perspective} and, from this, derived that the least we can do is approach questions from multiple angles.
In philosophy, the latter approach has found its name in \emph{epistemological pluralism} \cite{miller_epistemological_2008}. 
It is \emph{confession} and \emph{mandate} at once.
Confession---as it accepts that no single perspective can lead to complete knowledge.
Mandate---as it demands that multiple perspectives are considered simultaneously.
Epistemological pluralism has made its way into several quantitative and qualitative fields,
promising to render existing methods more robust and transparent and to yield more holistic insights.

In this work,
we develop the perspective of epistemological pluralism for \emph{network science}, yielding a framework we call \emph{\NP}.
Network science models \emph{complex systems} as networks and analyzes them using methods rooted in graph theory and statistical physics  \cite{barabasi_network_2016,torres_why_2021,newman_networks_2018}. 
The process of modeling a complex system as a network constitutes an act of perspective-taking, 
where one network only represents one possible perspective on a complex system. 
Hence, from an epistemological standpoint, 
multiple networks exist simultaneously, 
and any attempt to approach domain-driven questions should not rely on a single network only. 
Rather, several different perspectives should be considered in what we call a \emph{multi-network analysis}.

\NP is an inherently \emph{interdisciplinary} task.
In fact, it has been primarily motivated by an interdisciplinary attempt to analyze data from the legal domain using network-analytical tools, 
where choosing the most obvious network model and applying conventional network metrics
has---from a domain perspective---yielded incomplete or unimpressive results.
Only after considering multiple perspectives at once, the results became more meaningful.

\textbf{Contributions.}
In this work, we develop \NP as a conceptual framework and illustrate it via a multi-network analysis of the (German) legal system. 
We make four contributions---%
one concerning overarching conceptual questions~(C\ref{contr_np}) 
and three establishing novel ways of modeling (legal) citation networks~(C\ref{contr_legislation},~C\ref{contr_hierarchy},~C\ref{contr_document_types}).

\begin{compactenum}[(C1)]
	
	\item \textbf{Developing \emph{\NP} as a conceptual framework.}\label{contr_np}
	We reconsider network modeling and analysis in light of epistemological pluralism,
	developing \NP as a conceptual framework that leverages multi-perspectivity to yield more complete, meaningful, and robust results.
	Beyond its exemplary application to the legal system, this work intends to be a blueprint for any \emph{interdisciplinary} network-science endeavor.
	
	\item \textbf{Analyzing legislation based on its \emph{usage in jurisprudence}.}\label{contr_legislation} 
	Given that legal-network research has largely focused on jurisprudence (see \cref{sec:background_legal_network_science}), 
	we \emph{shift the focus} toward legislative rules (i.e., statutes and regulations),
	characterizing them substantively---beyond meta-level descriptions---through the lens of their application in judicial practice.

	\item \textbf{Integrating \emph{heterogeneous hierarchy} into citation networks.}\label{contr_hierarchy}
	Our analyses of the legal system show that both the way branches of government are organized and the internal structure of their documents exhibit \emph{heterogeneous hierarchy}.
	We move beyond current models that predominantly resolve citation networks to the document level by combining reference information with hierarchy \emph{above} and \emph{below} the document level.
	
	\item \textbf{Analyzing citation networks \emph{across document types}.}\label{contr_document_types}
	Studies of legal citation networks have often investigated citations to documents of the \emph{same type}, e.g., decisions citing other decisions (see \cref{sec:background_legal_network_science}).
	There exist, however, links \emph{between} and \emph{within} virtually all legal document types.
	As a step toward representing legal systems more comprehensively,
	we instantiate a network model that covers documents across different branches of government, namely jurisprudence and legislation,\footnote{This also covers the executive branch to the extent that it acts regulatorily, i.e., by issuing regulations.} 
	integrating the bipartite relation between the two document types with the unique metadata and hierarchical structure available for each side. 
	
\end{compactenum} 

\textbf{Structure.}\quad
We start by providing the relevant background on epistemological pluralism and legal network science (\cref{sec:background}).
Next, we develop the specifics of \NP (\cref{seq:network_pluralism}),
elaborating on how modeling choices and metric parameters relate to a space of perspectives
and why this relation may generate more complete and meaningful insights.
We proceed to illustrate our framework via a hands-on multi-network analysis of the legal domain (\cref{seq:NP_in_action}).
This includes presenting the dimensions of variation within legal systems in general (\cref{seq:legal_system_presentation}), 
leveraging these dimensions to define a network space for the German legal system (\cref{seq:legal_network_space}), 
and working with this space to explore specific questions about the relationship between court decisions on the one hand and statutory and regulatory documents on the other (\cref{seq:case_studies}).
We conclude by discussing our findings and sketching opportunities for future work on \NP and legal network science (\cref{seq:discussion}).

\section{Background}\label{sec:background}

Both \NP as a concept and our motivation to operationalize it are rooted in several fields,
ranging from \emph{philosophy} to general \emph{network science},  
to \emph{machine learning} and \emph{applied statistics}, 
to the specialized domain of \emph{legal networks}.
In this section, we first introduce the general idea of \emph{epistemological pluralism} and demonstrate how it---explicitly or implicitly---affects quantitative methodology.
We then illustrate why the current state of \emph{legal network science} has motivated us to develop a pluralist approach to analyzing \emph{legal systems} specifically and \emph{complex systems} more broadly.

\subsection{Epistemological Pluralism in Quantitative Research}\label{seq:epistemological_pluralism_quantitative_research}
As suggested by its Latin root, 
\emph{pluralism} in general denotes a multiplicity of views. 
A foundational concept, pluralism has been applied to and transformed by many domains.
Within epistemology, a branch of philosophy that investigates the process of reaching knowledge,
pluralism has led to the conclusion that any content is \emph{perspectival}, i.e., relative to a specific point of view \cite{redding_what_2003}. 
This means that no phenomenon can be ``fully explained by a single theory nor be investigated using a single approach'' \cite{kellert_introduction_2006}.
Instead, multiple approaches are required.

With varying motivations,
the mandate to include an epistemologically pluralistic approach has been included in interdisciplinary and quantitative research.
In \emph{machine learning}, engineers have adopted the practice of using a grid search
to identify a combination of hyperparameters that yields the best performance on a (predominantly predictive) downstream task, based on some evaluation metric(s) \cite{bergstra_random_2012}.
In pluralistic terms, this acknowledges a space of co-existing perspectives.
However, navigating this space is mostly framed as an \emph{optimization problem}, 
aiming to find either \emph{one} optimal perspective or a set of perspectives that \emph{together} yield the best performance through ensemble methods \cite{goos_ensemble_2000}.

From a \emph{statistical} perspective, researchers have suggested a pluralistic approach to \emph{increase robustness and transparency} through \emph{multiverse analysis} \cite{wayland_mapping_2024,ganslmeier_estimating_2025,steegen_increasing_2016}. 
Multiverse analysis aims to establish that a hypothesis holds across multiple model configurations (multiverse) instead of relying on only one single model (universe),
which might be an outlier produced by a specific series of modeling choices.
Implicitly, this assumes that a model (configuration) only presents one perspective and is unable to test a hypothesis (elucidate a phenomenon) completely.

Within \emph{network science}, the development of \emph{multilayer networks} has been motivated by the inability of simple graphs to fully capture the heterogeneity usually exhibited by complex systems \cite{kivela_multilayer_2014}.
This framework already embraces variation and can be interpreted as a technical response to an epistemologically identified problem.
While this line of research has formalized a network model that can capture \emph{more} information, 
it has not leveraged its added expressivity for interdisciplinary research through multi-network analysis. 
Moreover, while multiverse analysis has seen some adoption in network science \cite{coupette_all_2024}, 
the pluralistic perspective on network science has not yet been systematically developed. 

Building on these prior developments, 
\NP investigates how networks relate to each other \emph{as perspectives} and how considering a multiplicity of networks can yield novel insights.

\subsection{Legal Network Science}\label{sec:background_legal_network_science}

To gain a quantitative understanding of complex legal systems, 
researchers have leveraged the perspective and methods of network science (for detailed expositions, see \cite{coupette_juristische_2019,whalen_legal_2016}). 
Depending on their research interest, they have predominantly constructed \emph{social networks}, focusing on actors such as judges that work together \cite{katz_hustle_2010,tam_cho_legislative_2010,porter_network_2005}, 
or \emph{document networks}, focusing on how legal texts are interconnected by references and citations \cite{derlen_goodbye_2014,leicht_large-scale_2007,coupette_measuring_2021,katz_complex_2020,jeong2025centuryevolutioncomplexityunited}.
In the latter category, 
many works have investigated corpora of judicial decisions---%
seeking, for example, to identify important court decisions \cite{fowler_network_2007,derlen_goodbye_2014,van_opijnen_marc_citation_2012}.
Others have adopted a meta perspective and investigated the complexity of the legal system as such \cite{katz_complex_2020,ruhl_measuring_2015,coupette_measuring_2021}. 

Methodologically, researchers have recently experimented with more complex graph representations 
such as \emph{hypergraphs} \cite{coupette_legal_2023,matsumoto_hypergraph_2025}.
They have also started to combine structural representations with methods of natural language processing,
e.g., to extract implicit references \cite{sulis_exploiting_2022} 
and facilitate downstream tasks such as document similarity assessment \cite{bhattacharya_legal_2022}, statute identification \cite{paul_lesicin_2022}, or case law clustering \cite{mohammadi_combining_2025}.
Beyond the dominant scope of analyzing \emph{either} decisions \emph{or} statutes,
some studies have experimented with combining statutory and judicial documents 
\cite{bommarito_empirical_2011,beckedorf_komplex_2025}.

Although both judicial and legislative citation networks have been examined extensively using standard network-analysis methods, 
the substantive characterization of statutory units based on their appearance in jurisprudence 
via a combined, heterogeneous document network  
remains underexplored. 
Furthermore,
references are often resolved to the document level or---on the legislative side---to the section level.
While appearing \emph{natural},
this level of detail does not align with the more fine-grained reference granularity expected by (and from) legal professionals,
indicating a need to examine and reconsider assumptions regarding this specific modeling decision. 

Against this background, 
by applying \NP to the legal system, 
we systematically assess what perspectives different levels of resolution correspond to and how changing these perspectives affects the results of network metrics. 
In parallel, we further explore the relation between jurisprudence and legislation,
presenting a range of novel, substantively motivated research questions 
and demonstrating the value of multi-perspectivity in legal network science.

\section{\NP}\label{seq:network_pluralism}

In this section, we develop \emph{\NP} as a conceptual framework to implement epistemological pluralism in network science.
\NP is based on the understanding that every complex system allows multiple coexisting network representations. 
\emph{A priori}, all of these representations are equally valid, i.e.,
there is no single correct network model.
Rather, adopting multi-perspectivity can yield more complete, meaningful, and robust analytical results. 

\subsection{Networks are perspectives in a network space}

A key concept in epistemological pluralism is that of \emph{perspective}.
At a high level, a perspective can be defined as a point chosen to observe a target object \cite{redding_what_2003}.
In a \emph{perspective space}, any such point constitutes a \emph{direction} toward the observed object from a specific \emph{distance} and a unique \emph{angle}.
Beyond the location indicating from \emph{where} an object is observed,
there are several ways \emph{how} it is seen.
This can be compared to specific pairs of glasses that capture only some reflections of the observed object while omitting others.
All these aspects affect both \emph{sharpness} and \emph{range} of sight.

Any perspective generates a \emph{unique} view of an object,
and any object can be viewed from multiple perspectives.
Adopting several perspectives allows us to aggregate many \emph{isolated} observations into one more sophisticated image of the observed object.
As each perspective is limited, multi-perspectivity is key.
In a research context, the \emph{space of perspectives} is constructed from many domain-specific \emph{dimensions of variation}, 
and points of observation are defined by combinations of values along these dimensions.
Thus, any \emph{specific} perspective can be viewed as one element of the Cartesian product of the dimensions of variation.  
Researchers will choose points of observation and pairs of glasses based on their research interest and their intuition about the studied object.\footnote{%
	In quantitative legal studies, for example, pairs of glasses can be seen in different modalities of interpreting the law, e.g., law as \emph{networks}, law as \emph{text}, or law as \emph{code}.
}

In network science, the idea of multiple coexisting perspectives translates to the insight
that a network does not represent the complex system as such,
but rather constitutes a certain perspective on that system.
Many possible network representations coexist,
constituting what one might call a \emph{network space}.
Every network in this space is defined and accessed based on a set of modeling decisions (such as network type, node and edge sets, mode of observation, or any further modification and transformation).
Investigating the network space by considering multiple networks at once 
allows us to approach and explore research interests from complementing and contrasting domain perspectives.
It requires us to map the space of perspectives to the space of networks
by aligning the dimensions of variation found in a domain with network-modeling choices or network-metric parameters. 
This is a highly interdisciplinary task, and it is essential to \NP.

\subsection{Multi-network analysis enables new types of insights}\label{sec:NP_new_insights}

\NP enables \emph{multi-network analysis}---i.e., the combined analysis of multiple networks representing the same data. 
This approach offers several benefits for discovering, developing, and answering network-analytical research questions. 

When discovering and developing research questions, \NP increases flexibility and expressivity by moving the starting point of any research endeavor from a specific network to a more general network space.
Instead of constructing a single network early in the research process,
the network space is analyzed to explore the domain and potentially discover additional questions of interest.
This strengthens researchers' intuitions about the complex system under study
and allows them to uncover unknown patterns or unexpected biases at an early stage.
It also provides early feedback on the promise of existing research questions and their assumptions.

When aiming to answer pre-determined research questions,
\NP can generate more informative results
by \emph{complementing} and \emph{contrasting} perspectives---%
i.e., combining observations from several perspectives and deriving additional insights from the differences between these perspectives. 
Thus, \NP seeks to address some of the shortcomings of conventional network analysis at the \emph{modeling stage}, 
supplementing approaches that develop more expressive individual models and metrics. 
For example, based on domain-informed filtering functions,
subsystems of a complex system can be analyzed \emph{in isolation} to accommodate narrower research scopes, 
or they can be studied \emph{in conjunction} to compare perspectives and gain insights into the relationship between them. 

Beyond the modeling stage,
\NP aims to enhance our understanding of how certain perspectives influence specific outcomes 
when developing and applying analytical methods and metrics.
This can pinpoint potential biases,
reveal the need to reconsider or improve network metrics,
and enable more informed interpretation(s) overall.
For example,
metrics built on aggregation can be scrutinized by examining the composition of the processed units in relation to their influence on the aggregated result.
If an imbalance is identified, one can aggregate values that are normalized separately for each perspective. 
Furthermore, network metrics can combine observations from several perspectives,
e.g., from measurements taken at different levels of granularity.
Parameterizing the relation between two perspectives, 
for example via dynamic thresholds, 
grants access to a space \emph{in between} different perspectives, 
yielding an even more fine-grained space of networks.

Beyond facilitating more complete insights,
multi-network analysis can also help ensure more robust results.
Investigating a given question from many perspectives shows if a pattern can be observed from various angles or rather depends on specific modeling choices.
While the former suggests that the pattern is not an artifact of analytical choices,
the latter does not necessarily suggest the opposite---%
it is the very nature of perspectives that they grant a unique view on an object
and enable observations that may be impossible to perceive from other perspectives.
Nonetheless, understanding how observations depend on perspectives increases \emph{transparency} and opens room for further evaluation.

\section{\NP in Action}\label{seq:NP_in_action}

Having presented the concept of \NP in general,
we will now apply it to our domain of choice: the legal system. 
We will proceed in three steps, 
\begin{inparaenum}[(1)]
	\item introducing the legal domain as a \emph{complex} system,
	\item deriving some important dimensions of variation, and 
	\item leveraging the identified heterogeneity in a multi-network analysis.
\end{inparaenum}

\subsection{Multi-Perspectivity for Legal Systems}\label{seq:legal_system_presentation}

\begin{figure}[t]
\centering
\begin{subfigure}[b]{.24\textwidth}
\centering
\includegraphics[ width=\linewidth]{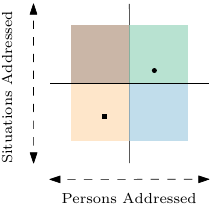}
\caption{Rules}\label{fig:rule}
\end{subfigure}
\begin{subfigure}[b]{.24\textwidth}
\centering
\includegraphics[ width=\linewidth]{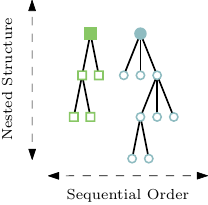}
\caption{Documents}\label{fig:documents}
\end{subfigure}
\begin{subfigure}[b]{.24\textwidth}
\centering
\includegraphics[ width=\linewidth]{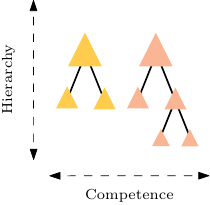}
\caption{Institutions}\label{fig:branches}
\end{subfigure}
\begin{subfigure}[b]{.24\textwidth}
\centering
\includegraphics[ width=\linewidth]{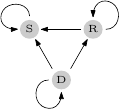}
\caption{References}\label{fig:interplay}
\end{subfigure}
\caption{\textbf{Law as a corpus of interconnected rules.}
	\emph{Rules} are the most granular unit of law, and they differ in their level of abstraction regarding the situations and persons addressed (\ref{fig:rule}). 
	\emph{Documents}, such as statutes, rulings, or contracts, are the carriers of rules (\ref{fig:documents}). 
	\emph{Institutions} in different branches of government (as well actors outside government) produce documents as one of their typical outputs (\ref{fig:branches}). 
	\emph{References} connect rules contained in documents of potentially different types, allowing them to build on and complement each other. 
	This is illustrated for the reference relations between statutes (S), regulations (R), and court decisions (D),   
	which together cover key parts of the legal system (\ref{fig:interplay}). 
   }\label{fig:legal_system}
  \end{figure}

Legal systems are built on text and act through people.
In different branches of government (and elsewhere), individuals or institutions create and interpret legal documents,
continuously producing new outputs that feed back into the system and combine into a growing corpus of rules, i.e., decisions at different levels of abstraction \cite{coupette_measuring_2021}.
This process might be bound to formal procedures, 
but the substantive changes to the law do not follow predetermined patterns.
Instead, these changes emerge from the complex interplay between texts, individuals, and institutions. 
This motivates the interpretation of legal systems as complex systems \cite{ruhl_harnessing_2017,katz_complex_2020,coupette_measuring_2021,vivo_complex_2025}. 
In \Cref{fig:legal_system}, we depict those components of legal systems that are crucial from the perspective of our study. 

\paragraph{Semantics of document structure and document authorship}

One role of the \emph{documented} legal system
is to \emph{index} and \emph{preserve} the actions of its agents.
Each branch of government (and non-government) fulfills a specific function,
and its actions manifest in documents of characteristic types (see Table~1 in \cite{coupette_measuring_2021}).
As such, documents are artifacts of a higher-level interaction between institutions.
Accounting for this interaction is important in any analysis,
and, as our case study will show, 
it unlocks a better understanding and contextualization of analytical results.

\begin{figure}
\centering
\begin{subfigure}[b]{.45\textwidth}
\centering
\includegraphics[ height=3.7cm]{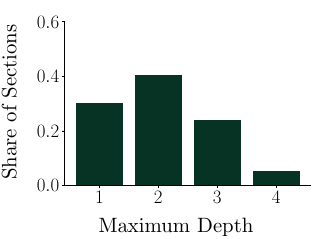}
\caption{Available Depth}\label{fig:legislation_depth}
\end{subfigure}
\begin{subfigure}[b]{.45\textwidth}
\centering
\includegraphics[ height=3.7cm]{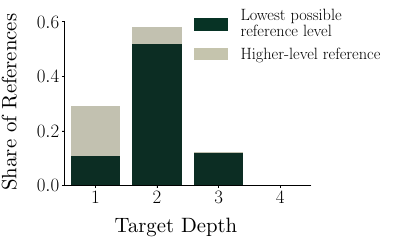}
\caption{Exploited Depth}\label{fig:reference_depth}
\end{subfigure}

\caption{\textbf{Reference granularity mirrors hierarchical structure.}
The legislation in our case study shows that most rules are not just distributed horizontally across sections but also distributed vertically across several levels of sub-sections (\ref{fig:legislation_depth}).
This elaborate substructure is reflected in the references made by judicial decisions, 
which refer to legislative units with a high level of detail, 
i.e., often below the section level, and mostly at the most granular level available (\ref{fig:reference_depth}). 
   }\label{fig:hierarchy_and_granularity}
  \end{figure}

As recognized in prior work \cite{coupette_measuring_2021,foerster_2022,jeong2025centuryevolutioncomplexityunited}, 
legal documents exhibit three core structural properties:
They are \emph{hierarchically structured},
consist of \emph{sequentially ordered} components,
and contain \emph{references} to other documents.
The first two properties enable a targeted identification of text components at a high level of detail.
As shown in \Cref{fig:hierarchy_and_granularity}, 
the data underlying our case study contains
extensive fine-grained structure within legislative documents (\ref{fig:legislation_depth}), 
which is also reflected in the references made by judicial decisions,
which tend to target legislation below the section level, where applicable (\ref{fig:reference_depth}).\footnote{%
	These results are not surprising. 
	Rather, they quantitatively reflect the level of detail expected from legal practitioners. 
	A normative indicator of the applicable standard can be found in reference and citation guidelines, e.g., from the Federal Administrative Court: 
	https://www.bverwg.de/rechtsprechung/urteile-beschluesse/zitierungen.
}
Neither the partitioning of documents by their \emph{producers} nor the access to them by their \emph{consumers} is random. 
Rather, both are inherently guided by legal semantics.

Branches of government are \emph{hierarchically} organized.
Here, hierarchy serves to allocate \emph{competency},
which can typically be broken down into a \emph{geographical}, a \emph{substantive}, and an \emph{instantiate} component.
Competency in practice does \emph{not} emerge from a dynamic interplay between institutions---%
rather, it results from a system that assigns cases to specific institutions or actors based on predefined rules (which may also include randomness).\footnote{%
	There are some exception to this.
  Under certain circumstances, parties can agree on a competent court (\emph{choice of forum clause}), or
  claimants can choose one of several competent courts.
  Further, there exist mechanisms to resolve uncertainty in cases that combine certain areas of law or that are otherwise heterogeneous.
  Such mechanisms often deem some higher-instance court competent to decide about competence.
}
While geographical and instantiate competency are of rather mechanical nature,
substantive competency encodes meaning into structure.
Laws establish the responsibility of specific courts for some broader area of law, 
and court-specific \emph{allocation plans} group certain types of cases and distribute them among their decision-making bodies (\emph{panels}).
These normative sources show that the bundling of cases is accompanied by substantive reasoning.

In summary, both document structure and document authorship go beyond a purely organizational function to encode legal semantics.
As such, they establish valuable dimensions of variation that can provide context and contribute to the definition of perspectives on legal systems.

\paragraph{Data model underlying our case study}

The heterogeneous information available about actors and documents 
results in a large space of perspectives.
For the purposes of our case study,
the space of perspectives contains different combinations of varying \emph{vertical} levels of granularity on each side
and \emph{horizontal} partitions to study and compare subsystems of a legal system.

To illustrate our approach, 
we instantiate a network representation of \emph{judicial}, \emph{legislative}, and \emph{regulatory} documents in Germany,
produced by their corresponding branches of government at the \emph{national} level.
From the multitude of potential \emph{reference relations} that exist within and between these branches (see \Cref{fig:interplay}),
we focus on references to legislative and regulatory units found in judicial documents.

Our data includes the entirety of the current German legislation and regulation in its consolidated form 
as well as a subset of court decisions issued by the highest federal courts.\footnote{%
The raw data for both parts of our corpus is provided by the \emph{Federal Ministry of Justice and Consumer Protection}.
While the substantive coverage of the statutory documents is extensive,
the coverage of documented court decisions starts only on January 1st, 2010.
Unfortunately, access to legal data in Germany is limited, e.g., it remains unclear why certain rulings are (or are not) (1)~published and (2)~made available in a machine-readable format.
More details on our raw data and preprocessing can be found in \Cref{appendix:data}.
}
We decompose all documents into their structural components
and reconstruct the institutional superstructure of the judiciary, 
which contains information about which panel of which court issued which specific decisions. 
Furthermore, we extract all explicit references from judicial documents toward any legislative unit indexed in our dataset.
This important preprocessing step builds upon the \emph{QuantLaw} package,
which offers a sophisticated, rule-based reference parser tailored to German legislation~\cite{quantlaw}.

\subsection{The Legal System as a Network Space}\label{seq:legal_network_space}

In the following, we sketch how to operationalize the structure and embodied semantics of our example system in a multi-network analysis.
This requires constructing a \emph{base network}
and linking its dimensions of variation to network-modeling decisions and network-metric parameters.
As a result, we obtain a network space that accommodates our perspectives of interest.

\paragraph{Constructing a base network}

The base network aims to capture all dimensions of variation of interest.
We call it a \emph{base} network because it is \emph{raw},
i.e., \emph{all} nodes and references are represented \emph{as observed}.
This network is neither focused on nor bound to a specific level of representational granularity (or \emph{resolution}).
To derive specific networks from this base network without losing information, 
further adjustments will be needed.

\begin{figure}[t]
\centering
\includegraphics[ width=0.95\textwidth]{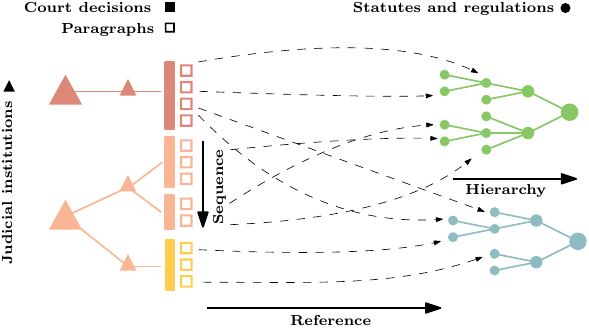}

\caption{\textbf{Base network.}~Our non-aggregated network representation includes hierarchical structure both within documents and within the institutions that produce~them.}\label{fig:raw_network}
\end{figure}

\Cref{fig:raw_network} depicts the model underlying our base network, 
combining two (directed) \emph{forests} with a directed \emph{bipartite graph}.
The forests represent branches of government, 
and their node sets represent either institutions and their components or documents and their components.
Institutional organization and document structure are reflected in different types of hierarchy edges, 
namely \emph{authorship edges}, \emph{organizational edges}, and \emph{containment edges}. 
The bipartite graph connects the forests
via \emph{reference edges}, 
with paragraphs of judicial decisions as sources and nodes in legislative trees as targets.
Table \ref{tab:network_description} provides an overview of our nodes and edges, along with their counts.

\begin{table}[tbp]
  \centering
    \begin{tabular}{lr}
    \toprule
           \bfseries Element Type & \bfseries Raw Count \\
    \midrule
           \underline{Actor Nodes}   & 141 \\
         - \emph{Courts} (J) & 8  \\
          - \emph{Panels} (J) & 133  \\

         \underline{Document Nodes}    & 3,709,831\\
           - \emph{Statutes} (L)  &  6,815 \\
           - \emph{Sections} (L) & 91,758 \\
           - \emph{Sub-sections} (L)  & 208,616  \\
           - \emph{Sub$^2$-sections} (L)  & 189,374 \\ 
           - \emph{Sub$^3$-sections} (L)  & 27,571  \\ 

            - \emph{Court Decisions} (J) & 74,214 \\
           - \emph{Text Paragraphs}  (J)  & 3,111,483\\
           \underline{Reference Edges} (J $\to$ L)&  1,685,078\\

    \bottomrule
    \end{tabular}%
      \captionsetup{width=\textwidth}

      \caption{\textbf{The base network consists of heterogeneous node and edge sets.}
      Nodes capture different kinds of actors, documents, and their respective components.
      Several types of edges introduce hierarchy (in organizations and document structure) and references between both sides, 
      where we group nodes by their function and omit the counts of authorship edges (= number of court decisions), organizational edges (= number of panels), and containment edges (= number of documents minus their highest-level counts) to reduce redundancy. 
      Letters in parentheses indicate the corresponding branch of government (\underline{J}udicial and \underline{L}egislative.)}

  \label{tab:network_description}
\end{table}

\paragraph{Deriving a multiplicity of networks}\label{seq:deriving_networks}

\begin{figure}[t] 
\centering
\vspace*{\fill} 
\begin{subfigure}[t][3cm][b]{.25\textwidth}
	\centering
    \includegraphics[ height=1.25cm]{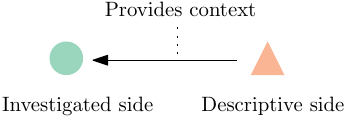}
    \caption{Focus}\label{fig:accessing_network_space:calibrate}
\end{subfigure}
\hspace{.5em}
\begin{subfigure}[t][3cm][b]{.3\textwidth}
	\centering
    \includegraphics[ height=2cm]{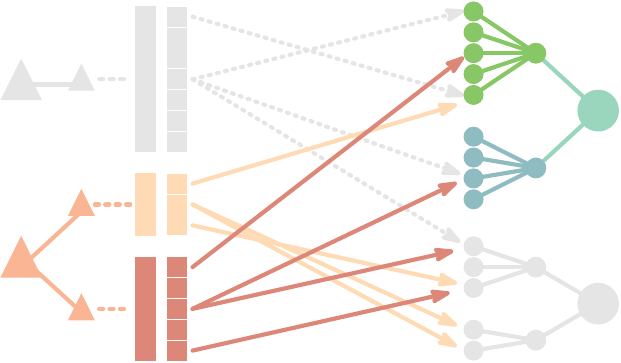}\quad
    \caption{Filter}\label{fig:accessing_network_space:filter}
\end{subfigure}
\hspace{.5em}
\begin{subfigure}[t][3cm][b]{.3\textwidth}
	\centering
  \includegraphics[ height=2cm]{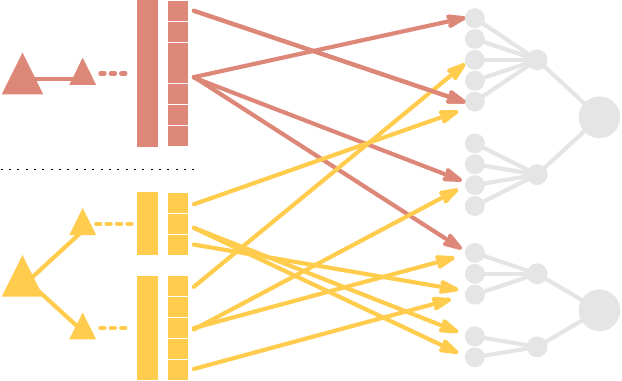} 
    \caption{Group}\label{fig:accessing_network_space:group}
\end{subfigure}

\vspace*{-1em}

\begin{subfigure}[b][3cm][b]{.3\textwidth}
	\centering
	\includegraphics[ height=1.25cm]{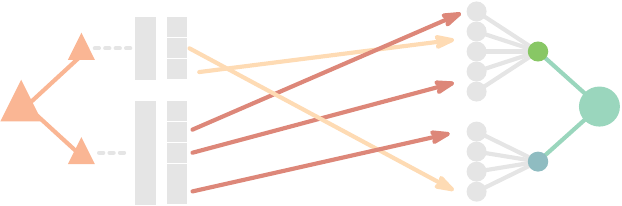}
	\caption{Aggregate}\label{fig:accessing_network_space:aggregate}
\end{subfigure}
\hspace{.5em}
\begin{subfigure}[b][3cm][b]{.3\textwidth}
	\centering
    \includegraphics[ height=1.25cm]{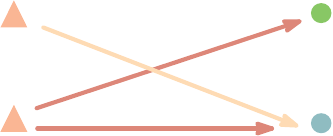}
    \caption{Bipartite Network}\label{fig:accessing_network_space:bipartite}
\end{subfigure}
\hspace{.5em}
\begin{subfigure}[b][3cm][b]{.3\textwidth}
	\centering
    \includegraphics[ height=1.25cm]{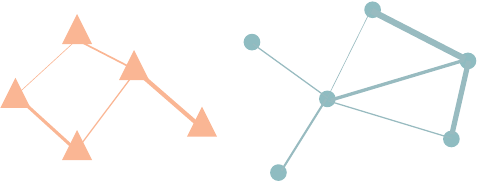}
    \caption{Projections}\label{fig:accessing_network_space:projections}
\end{subfigure}

\caption{\textbf{Accessing the network space requires domain-informed modeling decisions.}
First, we can \emph{focus} our bipartite relation on one of the two document sides, resulting in one investigated side and one descriptive side (\ref{fig:accessing_network_space:calibrate}).
This decision is often determined by a specific research interest.
Next, we can \emph{filter} (\ref{fig:accessing_network_space:filter}), 
\emph{group} (\ref{fig:accessing_network_space:group}),
or \emph{aggregate} (\ref{fig:accessing_network_space:aggregate}) the network to yield one or more bipartite networks tailored to our research question(s) (\ref{fig:accessing_network_space:bipartite}).
These choices depend on the substantive scope as well as the desired level of abstraction.
Optionally, we can further derive two \emph{projections} from the bipartite network (\ref{fig:accessing_network_space:projections}), capturing \emph{co-referencing} or \emph{bibliographic coupling}.
}
\label{fig:accessing_network_space}
\end{figure}

From the base network, we can derive a multiplicity of specific networks,
each reflecting a concrete perspective on the legal system.
At a high level, the modeling decisions that define each network
are choosing granularity levels (for each side),
(optionally) selecting a subset of each side to focus on, 
and defining what relation between units should be represented.
These general decisions are operationalized
through dynamic node selection,
filtering functions,
projections,
and edge computations that set thresholds,
leverage count data from different granularity levels,
and normalize values generated from different perspectives.
\Cref{fig:accessing_network_space} illustrates these operations.

Nodes are selected dynamically based on the chosen granularity, which corresponds to the level of abstraction.
In our case study, 
the granularity levels on the judicial side correspond to \emph{courts}, \emph{panels}, \emph{documents}, and \emph{paragraphs of documents} (\emph{Randnummern} in German).
On the legislative side, 
in decreasing order of abstraction, we consider \emph{statutes}, 
\emph{sections} (the only substructure guaranteed to be present in all statutes), 
and an arbitrary (and heterogeneous) amount of \emph{subsection levels} for each section.
This means that we could choose a network of courts referencing statutes, of paragraphs referencing sections at the first sub-section level, and so on.
Globally, the maximum depth of sections is observed in Sub$^3$-Sections,
which add up to five levels of total depth on the legislative side (including statute and section level).
Combined with four levels of depth on the judicial side, this results in $20$ potential networks. 

Each node set can be filtered using a number of \emph{filtering functions}. 
These can be based on some quantitative attributes, e.g., a minimum number of documents.
For example, on the judicial side,
we leverage an external and qualitative filter,
extracted from legislation and court-specific allocation plans,
that assigns specific types of cases to certain courts and panels.
For those networks that retain the relation of institutions to legislative units,
the filtering operation is relatively straightforward: 
Construct the judicial side at the document level, 
which requires leveraging the authorship relation between actors and documents,
and only select nodes that have been produced by the investigated authors.
Of course, a similar procedure can be applied to the legislative side,
e.g., by only considering sections from a specific subset of statutes that make up an externally determined \emph{area of law}.

A bipartite network can be projected onto each of its parts, 
resulting in a \emph{one-mode projection}.
Projecting a bipartite reference network onto judicial nodes allows comparing courts, panels, court decisions, or paragraphs among themselves; 
whereas projecting onto legislative nodes allows assessing, e.g., 
which sections are used similarly by the courts. 
This means that each of the $20$ bipartite representations of the legal system can yield two more projected networks,
resulting in a total of $20 + 40 = 60$ networks (without considering any filtering operations). 
While projections will often lose information, 
there are numerous ways to compute edge weights to minimize information loss.
One strategy preserves metadata about each co-occurrence in external tables or edge attributes, 
allowing researchers to analyze the composition of a co-occurrence relation (e.g., based on the source).
Alternatively, one could consider a variety of weighting functions in parallel
to derive more holistic insights by combining and contrasting a multiplicity of metrics. 
In the spirit of the latter approach, we leverage multi-perspectivity to construct and combine several weighting functions. 

While multiple relations between the same nodes can be represented as edge weights, computing these weights generally involves \emph{counting} relations. 
This counting, however, is relative to a perspective, and it can even combine perspectives. 
In a co-reference projection, for example, one could count the number of court decisions that contain references to two sections $i$ and $j$, or one could count the total number of such co-occurrences.
From a different perspective, one could also count the number of \emph{paragraphs} in which $i$ and $j$ both occur, 
which can yield very different results.
Perspectives can be combined by separating \emph{observing} from \emph{counting}, assigning each step its own perspective. 
For example, we can count all court decisions (more abstract unit) in which we observe at least $k$ co-references between two sections $i$ and $j$ at the (more granular) paragraph level.
When complementing different perspectives in this way,
the introduced threshold (e.g., the \emph{minimum number of paragraph-level co-references})
acts as a parameter, 
allowing a multiplicity of edges to coexist. 

To account for scale distortions,
edges weights are often normalized,
i.e., we relate values to some common scale with the objective of \emph{putting things into perspective}.
The specific scale used for normalization again encodes a certain perspective.
For example,
one could relate each reference to the total number of references in a specific decision, or to the decision length.
By partitioning the network 
and scaling values separately in each~part, 
normalization can also be used in conjunction with filters. 

Overall, 
heterogeneous (legal) systems can be examined through the lenses of many different networks, each resulting from a specific combination of modeling decisions. 

\paragraph{Resolving structural inconsistencies via dynamic counting}\label{seq:dynamic_counting}

While managing the bipartite network using basic \emph{aggregation} and \emph{filtering} operations seems straightforward, 
the base network contains two facets of heterogeneity that need to be addressed and resolved to avoid unwanted distortions.

First, the sections exhibit various \emph{levels of structural depth}.
While we can assume that there is a consistent relation between statutes and sections across the whole corpus,
the structure below the section level is inconsistent.
This means that if we chose a granularity below the section level,
a na\"ive approach could potentially disregard sections without subsections completely and inadvertently lose information.
Second, references vary in their \emph{level of structural detail}, 
i.e., judicial decisions may refer to the same norm at different levels.
Thus, we need to take extra care to retain information about all references that target units above or below the chosen level of resolution. 

Since \NP should change the perspective without affecting the underlying count data, 
we need an intermediate processing step to resolve the inconsistencies identified above.
This is a nontrivial task that, to the best of our knowledge, has not been systematically addressed in prior work. 
We propose a \emph{dynamic counting} algorithm to solve it. 
Roughly, our approach \emph{aggregates} references when moving to higher levels and \emph{distributes} references when moving to lower levels of abstraction.  
A~detailed description of our procedure can be found in \Cref{appendix:dynamic_counting}.

\subsection{Case Studies}\label{seq:case_studies}

Given the network space arising from complex legal systems as sketched above, 
we now demonstrate how to integrate a pluralist approach into domain-specific research questions and the methods designed to answer them. 
Our investigation focuses on legislative units (\emph{investigated side}), 
which we describe through the lens of institutions and documents from the judicial sphere (\emph{descriptive side}). 
Further domain-driven research could reverse these roles 
and view court decisions or judicial institutions through the lens of their legislative connections.
We offer two complementary case studies, 
focusing on the position and characteristics of individual sections (\ref{sec:case_study_nodes}) 
before examining how sections relate to one another based on their joint usage in jurisprudence (\ref{sec:case_study_edges}).
In both case studies,
we start from common research questions,
investigate them through multi-network analysis,
and explore how our findings could help refine the initial questions.

\subsubsection{Characteristics Beyond Centralities}\label{sec:case_study_nodes}

A classic research interest in many applied network-analysis studies lies in identifying the most important nodes of a network based on a range of \emph{centrality measures}.
However, such measures are only meaningful to the extent that they can be linked to some domain property that translates to importance as captured by the measure at hand. 
For legal citation networks,
the question of node importance has mainly been asked with the goal of identifying significant precedents. 
Following pioneering work on the jurisprudence of the United States Supreme Court \cite{fowler_network_2007},
researchers commonly approach this task under the assumption that a reasonable proxy for the significance of a legal document is the number of other documents citing that document \cite{derlen_goodbye_2014,alschner_growing_2018,soh_network_2019}.
Given the known challenges of the vanilla approach based on in-degree counts,
many studies have also looked toward other centrality measures,  
seeking to adapt them to the dynamics of legal systems \cite{van_opijnen_marc_citation_2012,koniaris_network_2018,whalen_purposes_2020}. 
We complement these efforts by exploring how leveraging multi-perspectivity can further the goal of generating more meaningful insights,
thus emphasizing modeling decisions over metric~design. 

Like prior work, we start from the intuition that importance can, to some extent, be captured by citation frequency---or \emph{reference frequency} when denoting judicial mentions of statutes and regulations.
Beginning with the na\"ive baseline, we rank sections based on their total reference count, 
i.e., the number of times they are mentioned in any judicial decision in our corpus. 
From a network perspective, 
this corresponds to the number of incoming reference edges,  
where one decision (and even one paragraph of a decision) can contribute more than one edge. 
To be precise, we consider the \emph{in-degree} of section nodes in the directed bipartite graph contained in our base network. 

\begin{figure}[t]
	\centering
	\includegraphics[ width=0.65\textwidth]{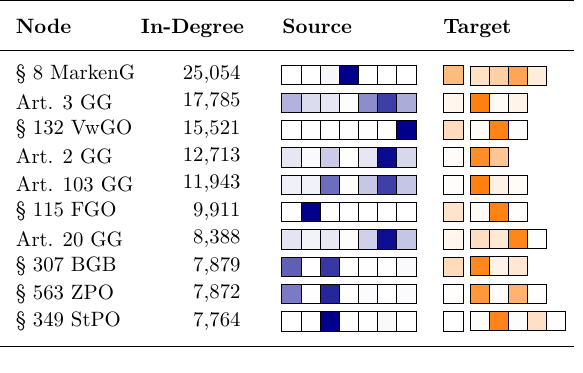}
	\caption{\textbf{Analyzing edges sets as distributions adds context}.
		\emph{Source distributions} (grouped by courts) show that \emph{overall} importance tends to be driven by a specific subset of courts.
		\emph{Target distributions} indicate that courts predominantly refer to specific subsections of prominent norms. 
		For the source distributions, each court is represented by a box. 
		The box coloring depicts the court-specific section reference frequency, scaled by the court's total number of references. 
		For the target distributions, the first, slightly separated box represents the section, and all subsequent boxes denote subsections.
		Here, the coloring corresponds to the share of \emph{direct} references, highlighting that there is no uniform distribution of references among subsections.
		Statute abbreviations are resolved in \Cref{appendix:abbreviations}.
	}
	\label{fig:in_degree_and_distributions}
\end{figure}

The two leftmost columns in \Cref{fig:in_degree_and_distributions} list the fifteen most frequently referenced sections along with their in-degrees.
However, assuming that the in-degree ranking reflects the importance of these sections in a legally meaningful sense would be misguided. 
Most of the listed sections are procedural, 
and while procedural norms are crucial for a decision to be issued,
they are often substantially irrelevant to the main problem(s) of the case.
Moreover, the sections come from very different areas of laws.
As very general norms within their respective fields,
they might share some \emph{functional} similarity. 
Beyond this, however, there is little reason to compare their importance,
as their legal environments (and the processes leading to their occurrence in court decisions) are too distinct to derive meaningful insights from a comparison.

Scrutinizing these preliminary results already highlights
that ``importance'' is a multi-faceted and relative concept that cannot be fully captured by the in-degree, 
which only provides a vague and noisy indicator of what constitutes important sections. 
To contextualize both the underlying question of importance and the metrics to answer it,
we propose to combine any metric with structural information above and below the document level,  
and to explore different metric parameters to assess their robustness and explanatory boundaries.

\paragraph{Adding context by treating incoming edges as distributions}

Our baseline setup aggregated the existence of references into one simple count,
disregarding the structural context of both reference source and reference target.
Instead of reducing the incoming edges to a single metric,
they can be treated and analyzed as a categorical distribution,
where the categories reflect some relevant property of either the descriptive or the investigated side. 
This can be considered a \emph{grouping} operation.  

On the descriptive side, we can classify references by authorship to derive a \emph{source distribution}.
This reveals how the relevance of a section depends on the perspective---here, the perspective of any given court.
Classifying references by authorship elucidates by which courts (or panels) a section is cited often.
When combined with information about the \emph{substantive competence} of each court,
the resulting distribution can be leveraged to approximate the area(s) to  which a specific section belongs.
This can be made even more specific by individually normalizing the perspective of each court, i.e., rescaling the references of a court $c_i$ toward a section $s_j$ by $c_i$'s total number of references.
The source distribution of the ten most frequently referenced sections (as identified by the general in-degree) is depicted in the third column of \Cref{fig:in_degree_and_distributions}. 
From these profiles, we see that many of the referenced sections are almost exclusively referenced by one or two courts, 
while others are spread out more evenly, 
especially \emph{articles}\footnote{%
	Articles in the German constitution are structurally equivalent to sections in other German laws. 
} 
of the constitution (German \emph{Grundgesetz}) that are intended to bind \emph{all} areas of law.
In other words, the \emph{importance} of a node depends on its belonging to a specific part of the complex legal system, 
and it can be analyzed either by focusing on one such a part or by perusing the composition of aggregate metrics.

On the investigated side, the \emph{target distribution} shows how references are spread vertically and horizontally across the \emph{substructure} of a section, 
exposing if an importance measure relates to the section \emph{as such} or is rather driven by the prominence of one or more subsection(s).
Obtaining this information requires considering multiple levels of granularity simultaneously. 
We depict the profiles of the ten most referenced sections in the rightmost column of \Cref{fig:in_degree_and_distributions}. 
Within the chosen subset of sections,
we can observe a recurring imbalance within the target distribution:
Vertically, subsections are referenced more often than the sections containing them, and horizontally, profiles are often dominated by one or more subsection(s). 

\paragraph{Exploring metric sensitivity to modeling decisions via multi-perspectivity} 

Individual metrics, such as the in-degree considered above, 
often reflect a specific perspective that is not guaranteed to reveal a persistent pattern. 
To evaluate metric uncertainty and obtain more comprehensive results, 
multiple perspectives ought to be combined. 
In the baseline setup, we counted overall references, 
but these are not the only units to count. 
We can, for example, instead count the number of \emph{decisions} that cite a section at least once or---to parameterize---at least $k$ times.

\begin{figure}
    \centering
    \begin{subfigure}[b]{0.495\textwidth}
        \centering
	\includegraphics[ width=.96\textwidth]{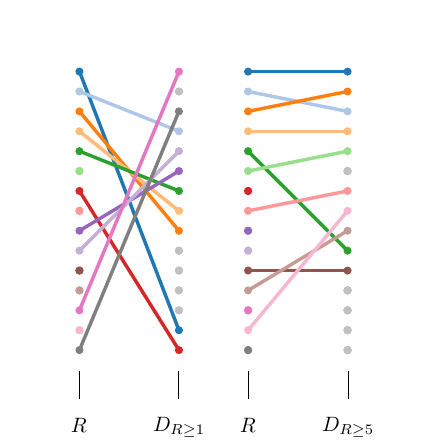}
        \caption{Pairwise Modeling Comparison}\label{fig:pairwise_comparison}
    \end{subfigure}
    \hfill
    \begin{subfigure}[b]{0.495\textwidth}
        \centering
	\includegraphics[ width=.9\textwidth]{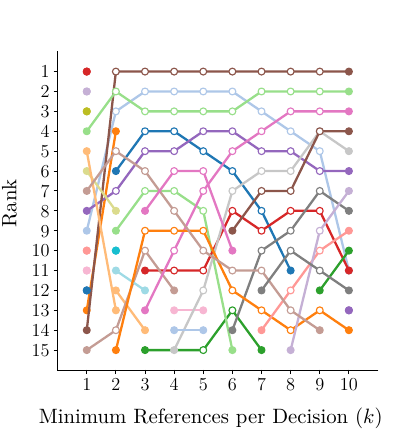}
        \caption{Minimum Reference Sensitivity}\label{fig:sensitivity}
    \end{subfigure}
   \caption{ \textbf{Importance ranks are sensitive to modeling and parameter choices.}
	Ranking sections ($y$) not by the absolute reference count ($R$) but by the number of decisions that refer to them at least $k$ times ($D_{R\geq k}$) reveals some overall sensitivity to a changing $k$ (\ref{fig:pairwise_comparison}). 
	Two thirds of the most referenced norms at $k=1$ are not among the top-referenced norms for $k\geq 2$ (\ref{fig:sensitivity}). 
	This pattern highlights the procedural character of many norms that are referenced mandatorily, yet lack substantive engagement. 
	The baseline metric (= overall in-degree) only captures a top layer of nodes that---without further specification---cannot be mapped to a \emph{general}, domain-aware notion of importance. 
}\label{fig:ranks}
\end{figure}

Computing such a metric over different values of $k$, 
i.e., different minimum numbers of references per decision, 
yields a series of ranks as depicted in \Cref{fig:sensitivity}.
These can be contrasted with the overall reference count depicted in  \Cref{fig:pairwise_comparison}.
We immediately see that our parameters, 
\emph{counting unit} and \emph{counting condition}, 
strongly affect both the selection of nodes and their ordering. 
For example, some sections recur across modeling conditions, 
but these sections are predominantly referenced at most once or twice per decision, 
which again suggests that they are very general or procedural in character.
While this does not mean that the sections in question lack importance, 
it underscores that importance must be evaluated, at the very least, relative to the function of a section. 
Consequently, it cannot be fully captured by one metric alone.

\subsubsection{Relevance Beyond Frequency}\label{sec:case_study_edges}

Beyond the characterization of individual norms, 
we can ask: Given a specific (seed) section, 
what \emph{other} sections should be considered of interest?
From a network perspective, this question concerns the neighbors of a node in a one-mode projection,
which requires us to \emph{somehow} project the bipartite relation from our base network (i.e., rulings $\leftrightarrow$ sections) onto the legislative side.
While the potential node sets of such a projection are clearly defined (nodes represent legislative units at a chosen depth $d$), 
to compute edges and edge weights, we need to take several decisions.
In the following, we assess how considering multiple perspectives can support these decisions, 
and how \NP can aid the development of metrics to describe and quantify the strength of relationships between two sections.

One plausible assumption underlying all co-occurrence analyses can be stated as follows:
Nodes that \emph{often} occur \emph{together} are \emph{relevant} to each other. 
In our context, this begs three questions:
\begin{inparaenum}[(1)]
\item When are two references considered to be referenced \emph{together}, i.e., how do we understand proximity in networks where nodes are part of a nested hierarchical structure?
\item What does \emph{often} mean, i.e., how can we operationalize the existence and strength of a co-reference relationship?
\item How can we define and dissect the concept of \emph{relevance} in the context of co-references?
\end{inparaenum}
In the following subsections, we consider each question in turn. 

\paragraph{Resolution  impacts relevance}\label{sec:proximity}

The term \emph{together} can be based on several notions of \emph{proximity}.
Viewing court decisions as texts, 
we could leverage techniques from natural language processing,
e.g., we could define proximity as co-occurrence within the same sentence or within some window of tokens. 
While this could potentially yield very fine-grained measurements, 
it risks discarding the valuable domain-specific information encoded in the \emph{structure} of court decisions---%
i.e., their chunking into paragraphs. 
In line with the representation chosen for our base network, 
we therefore opt to compute proximity based on structural components.

Given the multi-perspectivity enabled by \NP, 
even from a structural viewpoint, 
co-occurrence can be defined in several ways.
For example, we could count a co-reference if two sections are referred to by the same court, the same panel, the same decision, \emph{or} the same paragraph of a decision.
Thus, proximity depends on the level of granularity chosen for the descriptive side,  
and each definition of proximity corresponds to a certain perspective that represents interconnections from a unique angle.
Choosing either a single perspective or a combination of them yields different results.
In the following, we compare and combine co-reference observations at the \emph{decision} level and at the \emph{paragraph} level.

\begin{figure}[htp]
    \centering
    
    \begin{subfigure}{0.495\textwidth}
        \centering
        \includegraphics[height=4.25cm]{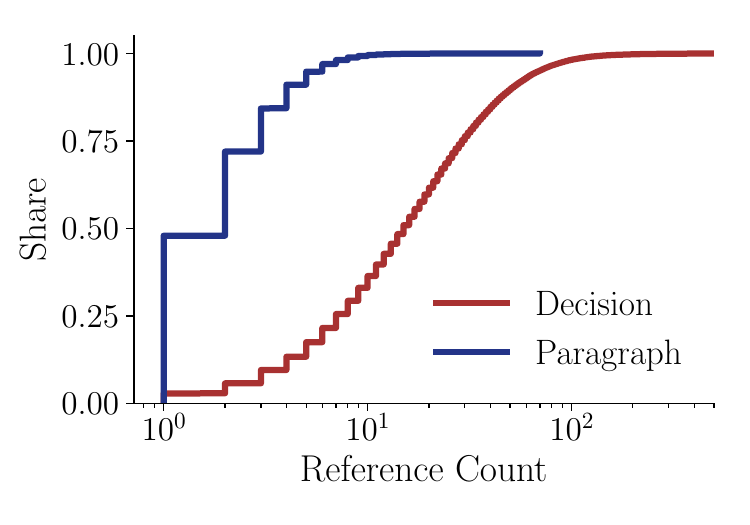}
        \caption{Outgoing References}
        \label{fig:outgoing}
    \end{subfigure}
    \hfill
    \begin{subfigure}{0.495\textwidth}
        \centering
        \includegraphics[height=4.25cm]{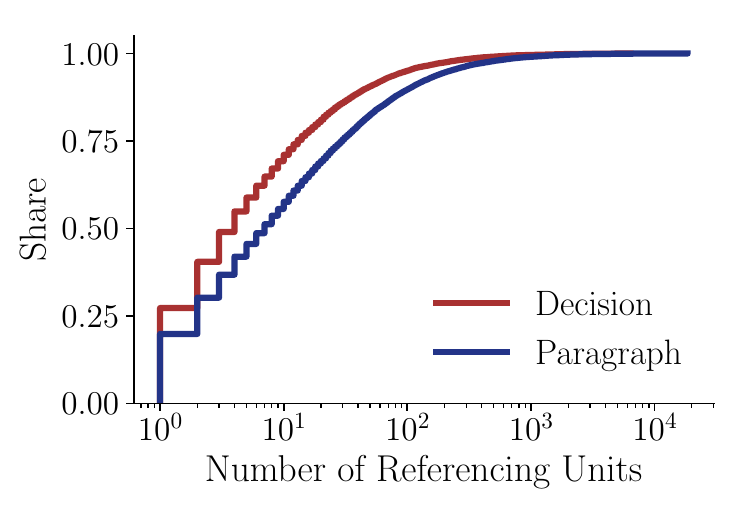}
        \caption{Incoming References}
        \label{fig:incoming}
    \end{subfigure}

    \caption{\textbf{Document structure encodes several notions of proximity.}
    The distribution of the number of references per referencing unit changes significantly as we move from court decisions to paragraphs of decisions (\ref{fig:outgoing}).
    Yet, the distribution of incoming references per section (measured in the number of referencing units) remains similar across both granularity levels (\ref{fig:incoming}).
    Thus, each level of granularity describes the same data in a distinct way, which must be accounted for in the interpretation of results. 
    }
    \label{fig:distributions_docs_and_paras}
\end{figure}

On average, a decision contains $22.7$ references.
From a legal perspective,
it is evident that a court decision can cover multiple questions and contain ``boilerplate references'' (e.g., asserting a court's jurisdiction over a case),
which suggests that measuring co-references at the \emph{decision level} is rather uninformative.
This domain-informed assumption can be quantitatively evaluated by analyzing the structural characteristics of the corresponding co-reference network.
Large-scale statistics on the distribution of outgoing references per decision or paragraph and the corresponding distribution of incoming references per section are illustrated in \Cref{fig:distributions_docs_and_paras}.

\begin{figure}
    \centering
    \begin{subfigure}[b]{0.45\textwidth}
        \centering
        \includegraphics[ width=\textwidth]{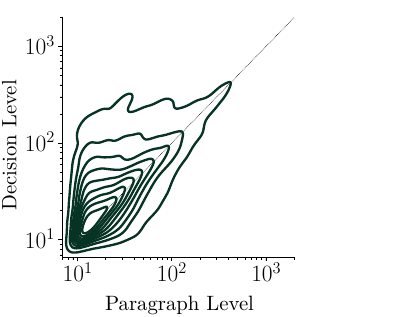} 
        \caption{All Sections}\label{fig:para-vs-dec-complete}
    \end{subfigure}
    \hfill
      \begin{subfigure}[b]{0.45\textwidth}
        \centering
        \includegraphics[ width=\textwidth]{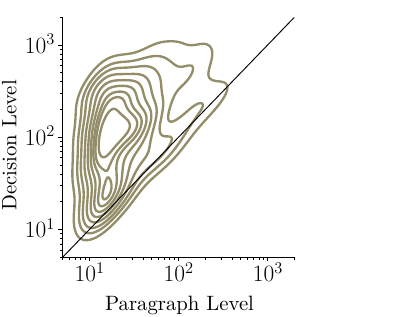} 
        \caption{Subset of Overrepresented Sections}\label{fig:para-vs-dec-subset}
    \end{subfigure}
    
    \caption{\textbf{Granularity levels yield different insights.}
    Both panels depict the distribution of co-reference counts of section pairs at the paragraph level ($x$) vs. the decision level ($y$) as a kernel density estimate.
    This grants insights into how the different levels of granularity relate to each other.
    Panel~(a) shows the distribution of all sections, 
    whereas panel~(b) shows the distribution of all pairs containing one of the sections identified as overrepresented at the decision level in \Cref{table:overrepresentation}.
    Comparing both distributions reveals how \emph{procedural norms} tend to be co-referenced more on the decision level and less on the paragraph level. 
}
\label{fig:overrepresentation}
\end{figure}

\begin{figure}
	\centering
	\includegraphics[ width=0.9\textwidth]{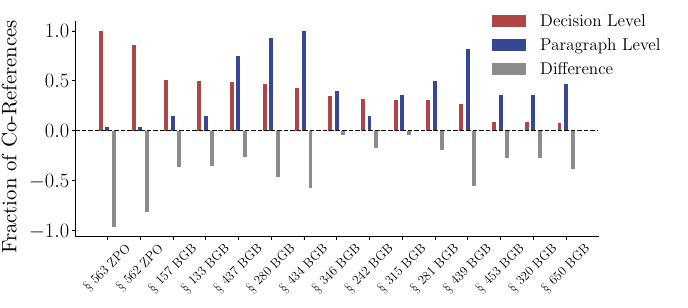}
\caption{\textbf{Top neighbors depend on network granularity.} 
Bars show the max-normalized co-occurrence share of the investigated section $\S\thinspace 433$~BGB (\textit{Purchase Agreement}), 
contrasting decision-level (coarse) resolution and paragraph-level (fine) resolution. 
Sections along the $x$-axis represent the union of the top ten norms across both resolutions. 
While procedural sections ($\S\S\thinspace 563$, $562$ ZPO) rank high at the decision level, they play a negligible role at the paragraph level.
Statute abbreviations are resolved in \Cref{appendix:abbreviations}.
}
\label{fig:bar_combarison_granularity_433}
\end{figure}

As depicted in \Cref{table:overrepresentation}, we find that across all sections, 
some sections are \emph{overrepresented} as close neighbors of other sections.
These sections are predominantly procedural norms, 
which---as already stated---are regularly referred to regardless of the specific content of the decision.
This pattern is also evident from \Cref{fig:overrepresentation},
which illustrates that the dominant norms tend to be co-referenced predominantly on the decision level, rather than on the paragraph level.
A specific example of the top fifteen neighbors of $\S\thinspace 433$~BGB (\emph{Purchase contract}) is given in \Cref{fig:bar_combarison_granularity_433},
which suggests that two procedural norms are most relevant to a purchase contract.
Indeed, these norms are the most frequently co-referenced sections at the decision level,
but they only establish the procedural sphere without touching the substantive core of sales law.
Given that our current one-mode projection is based on court decisions,
it is reassuring to find quantitative traces of procedural norms acting as formal prerequisites for any substantive question to follow.
However, crucial yet auxiliary relations between two sections do not translate to substantive relevance.
While this does not render the one-mode projection at the decision level obsolete, 
it points toward a need to include additional perspectives.

\begin{table}
	\centering

     \begin{tabular}{lrrp{0.6125\linewidth}}
\toprule
 & \multicolumn{2}{c}{\textbf{Share in \%}} &\\
\textbf{Section} & \multicolumn{1}{c}{\textbf{D}} & \multicolumn{1}{c}{\textbf{P}} & \textbf{[Unofficial] Section Name}\\
\midrule
§ 154 VwGO & 19.62 & 0.29 & [Costs of procedure]\\
§ 563 ZPO & 14.88 & 0.11 &Referral to a court of lower instance; decision on the merits ... \\
Art. 3 GG & 14.16 & 1.96 &[Equality before the law]\\
§ 126 FGO & 12.94 & 0.26 &[Decision procedure]\\
§ 97 ZPO & 12.07 & 0.13 &Costs of appellate remedies\\
§ 137 VwGO & 11.57 & 0.68 &[Grounds of appeal]\\
§ 562 ZPO & 10.41 & 0.06 &Reversal of the contested judgment\\
§ 135 FGO & 10.10 & 0.06 &[Costs of procedure]\\
§ 52 GKG & 9.58 & 0.25 &[Proceedings before administrative, tax, and social courts]\\
Art. 2 GG & 9.43 & 1.37 &[Personal freedoms]\\
\bottomrule
\end{tabular}
\captionsetup{width=\textwidth}

\caption{\textbf{Coarse granularity overemphasizes general and procedural norms.}
Ranking sections based on how often they account for \emph{at least one fourth} of all co-reference observations from other sections (as a fraction of all sections),
we see that tallying co-references at the decision level (D) yields a coarse-grained description,
emphasizing highly general or procedural norms 
that should not be weighted strongly from a substantively motivated, domain-driven perspective. 
Examining co-reference shares at the fine-grained paragraph level (P) reduces the noise significantly.
Statute abbreviations are resolved in \Cref{appendix:abbreviations}. 
}
\label{table:overrepresentation}
\end{table}

A more fine-grained perspective records a co-reference of two sections if these sections are referenced within the same paragraph.
Doing so avoids the noise introduced by norms that are generally important at a decision level.
Both in the given example (\Cref{fig:bar_combarison_granularity_433}) 
and at large (column ``Share in \% $\rightarrow$ P'' in \Cref{table:overrepresentation}), 
the measured relevance of procedural norms,
which dominates at the decision level,
is reduced to a negligible fraction.
Beneath the layer of general norms, substantively close sections are disclosed.
However,
the varying number of paragraphs and references per decision may introduce another type of bias, 
whereby a small number of exceptionally long decisions that repeatedly mention the same sections will increase the overall co-reference score: 
A high reference score due to a \emph{one-time}, in-depth relationship cannot be distinguished from an equally high score that is based on repeated co-references \emph{over time}.

\begin{figure}[t!]
	\centering
	\includegraphics[ width=0.7\textwidth]{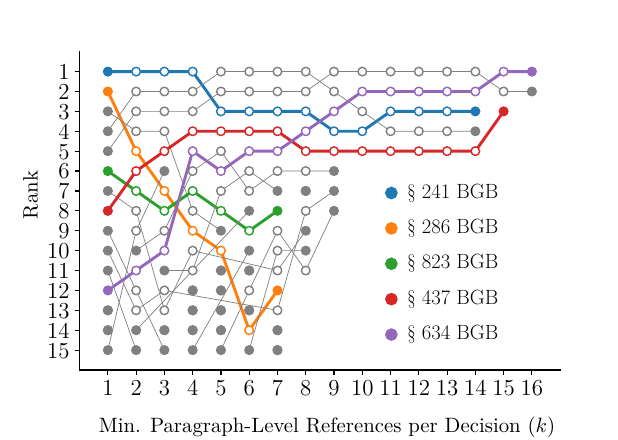}
	\caption{\textbf{Strength can be parameterized via thresholds in combined metrics.} 
		We compare the ranks of closest neighbors based on the \emph{number of decisions} in which the investigated norm ($\S\thinspace 280$ BGB---\emph{Damages for breach of duty}) and the named norm are referenced at least $k$ times on the paragraph level. 
        Statute abbreviations are resolved in \Cref{appendix:abbreviations}.
	}
	\label{fig:bar_comparison_granularity}
\end{figure}

To tackle the shortcomings of both levels of granularity,
we can consider a third metric,  
which aims to separate \emph{observing} a reference from \emph{counting} it, 
thus assigning each step its own perspective.
Specifically, we can count the number of \emph{decisions}
in which we observe \emph{at least one} (or at least $k$) co-references of sections $i$ and $j$ at the \emph{paragraph level}. 
Comparing the resulting rankings reveals that there is no universal dependency pattern between the relative importance 
and the paragraph-level threshold for counting co-references at the decision level (\Cref{fig:bar_comparison_granularity})---%
some norms move up in the rankings as $k$ increases, 
some move down, and others retain their position.
These trends add important context for investigating the relationship between two sections:
A larger $k$ does not necessarily correspond to higher importance;
rather, it allows researchers to explore and interpret the space between two complementary perspectives. 

In a nutshell, each level of granularity provides a distinct perspective on the same relationship.
\emph{Complementing}, \emph{contrasting}, and \emph{combining} multiple perspectives \emph{within} a metric yields more nuanced results.

\paragraph{Reciprocity complements relevance}

Moving beyond the definition of proximity, 
our second question asks about the strength of a co-reference relationship, 
i.e., what \emph{often} means. 
Starting from the intuition that one norm can be more important to another norm than vice versa, 
here, we explore the \emph{reciprocity} of the relationships between norms as exposed by normalization. 

\begin{figure}[t]
\centering
\begin{subfigure}[b]{0.45\textwidth}
    \centering
        \includegraphics[height=3.8cm]{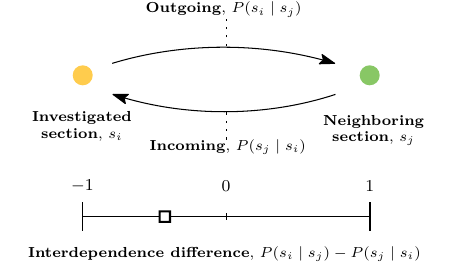}
        \vspace{.1em}
        \caption{Concept}\label{fig:bidirectionality:concept}
        \label{fig:pairwise_comparison}
\end{subfigure}
\hfill
\begin{subfigure}[b]{0.45\textwidth}
    \centering
        \centering
        \includegraphics[ height=4.25cm]{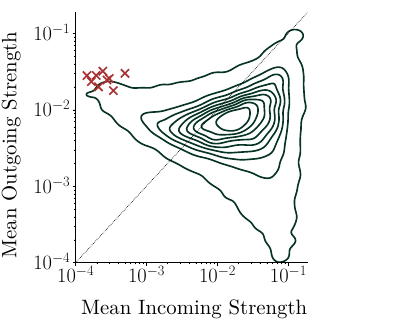}
        \vspace{1em}
        \caption{Meso-level patterns}\label{fig:bidirectionality:patterns}
        \label{fig:sensitivity}
\end{subfigure}
   
    \caption{\textbf{Bidirectionality contextualizes relations and characterizes sections.}
To investigate the bidirectional nature of relative co-reference metrics (\ref{fig:pairwise_comparison}), 
we contrast the mean incoming and outgoing co-reference strengths for all sections in a kernel density estimate (\ref{fig:sensitivity}). 
The `$\times$' markers in the right panel represent sections identified as \emph{dominant} in \Cref{table:overrepresentation}. 
Based on this concept, the dominant norms are clearly more important to other norms than vice versa.
}
\label{fig:bidirectionality}
\end{figure}

Any count metric relating an investigated section to its neighbors can be normalized by its sum over all neighbors (\emph{max normalization}), 
and the same can be done for the metrics of all neighbors. 
As a result, 
any pairwise relation can be interpreted in light of two opposing perspectives:
One direction evaluates how important a neighbor is to the investigated section, 
and the reverse direction captures the relevance of the investigated section to its neighbor.
We refer to these two measures as \emph{incoming} and \emph{outgoing} strength.
From a probabilistic perspective,
such bidirectional metrics for 
an investigated section $s_i$ and any of its neighbors $s_j$
encode the probability of observing $s_j$ given $s_i$, $P(s_j\mid s_i)$ (\emph{incoming}),
and the probability of observing $s_i$ given $s_j$, $P(s_i\mid s_j)$ (\emph{outgoing}). 
\Cref{fig:bidirectionality:concept} illustrates the concept. 
Accounting for multiple perspectives via bidirectionality may
provide fruitful insights into relationship imbalances and---at scale---the characteristics of individual nodes.
While presenting the full breadth of potential transformations to leverage these insights, 
e.g., by computing differences or (log-)ratios, 
requires its own considerations that are beyond the scope of this paper,
in \Cref{fig:bidirectionality:patterns}, 
we show how the mean \emph{incoming} and \emph{outgoing} strengths per section are distributed across a projection that is \emph{described} at the \emph{decision} level. 
At a high level, the variance of the distribution
suggests heterogeneous relations between sections
and points toward a latent hierarchy underlying the nodes.
If, for example, a section tends to be more important to its neighbors on average than vice versa,
this can again indicate its substantive generality or procedural nature.
This intuition is corroborated by the scatter markers in \Cref{fig:bidirectionality:patterns},
which represent the sections identified as dominant in \Cref{table:overrepresentation} and concentrate in a part of the distribution that emphasizes outgoing over incoming strength. 

\paragraph{Context refines relevance}

\begin{figure}
	\centering
	\begin{subfigure}{\textwidth}
        \centering
        \includegraphics[ width=\textwidth]{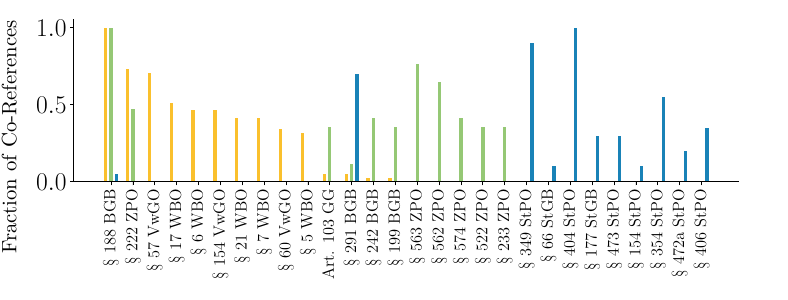}
        \caption{Normalized per Court}
        \label{fig:coocs_local}
    \end{subfigure}

    \begin{subfigure}{\textwidth}
        \centering
        \includegraphics[ width=\textwidth]{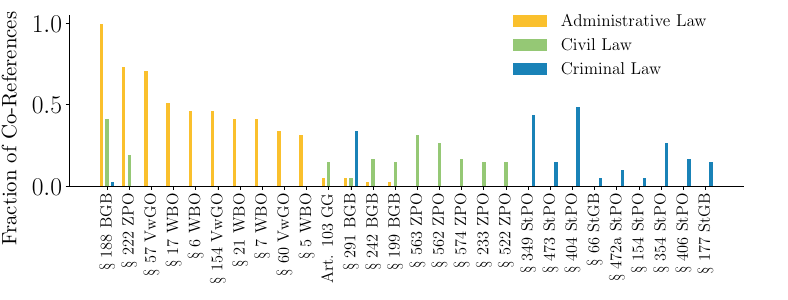}
        \caption{Normalized across all Courts}
        \label{fig:coocs_global}
    \end{subfigure}
\caption{\textbf{Top neighbors depend on the area of law.} 
	We show the share of co-references at the paragraph level, 
    max-normalized per court (\ref{fig:coocs_local})
    and max-normalized across all (filtered) co-occurrences in (\ref{fig:coocs_global}),
	for each section in the union of sections most relevant to $\S\thinspace 187$~BGB (\emph{Ending of legal deadlines}), 
	based on the references from three different panel groups (civil and criminal panels at the Federal Court of Justice as well as panels at the Federal Administrative Court). 
    Statute abbreviations are resolved in \Cref{appendix:abbreviations}.
}
\label{fig:bar_combarison_granularity_187}
\end{figure}

To identify complementing sections and distinguish different facets of relevance,
legal professionals work qualitatively,
applying different interpretive methods to understand the broader context of a section. 
We can supplement such assessments by 
applying our findings about source distributions (see \Cref{fig:in_degree_and_distributions} and accompanying text) to projected networks,
thus adding context to pairwise relations. 
Mapping the referencing panels and courts to areas of law 
can show which role a section plays in different legal contexts.
For example, 
if we consider $\S\thinspace 187$~BGB, a section that regulates the ending of legal deadlines within \emph{civil} law 
and group co-references by court of origin, 
as depicted in \Cref{fig:bar_combarison_granularity_187}, 
we see that this section is used not only in the civil context but also in administrative law and penal law.
A pattern like this, i.e., strong and unique relations across many areas of law,
suggests some generality, 
likely due to the highly functional character of the section.
Comparing the respective co-reference relations across the different subsets,
we observe that the investigated section is relatively co-referenced more in the decisions by administrative courts.
Since $\S\thinspace 187$~BGB is located in the civil code, 
this might seem counterintuitive. 
However,
it also highlights the reality that many procedures before the administrative courts touch upon questions regarding deadlines,  
illustrating how functional rules may be detached from substantive specifics.
In summary, grouping references by source can help identify relevant \emph{other} sections based on the respective area of law.

\section{Discussion}\label{seq:discussion}

Starting from the epistemological necessity of multi-perspectivity, 
in this work, we introduced \emph{\NP} as a conceptual framework to facilitate more complete, meaningful, and robust analyses. 
In our case study on the German legal system, 
we saw how considering several network representations of the same data simultaneously via \emph{multi-network analysis} can 
reveal dependencies between modeling decisions and analytical results 
or enable the construction of novel metrics that integrate insights from several different perspectives in \emph{network space}. 
In the following, we discuss the implications, limitations, and possible continuations of our work, first for \NP in general, then for legal network science specifically. 

\subsection{\NP}

\textbf{Implications.}\quad\NP diagnosed imprecision in network models, metrics, and research questions.  
At the same time, it provided a process for addressing these shortcomings: 
By considering multiple perspectives at once, 
\NP can add meaningful context to otherwise high-level findings,
produce more transparent and robust results,
and yield novel or more precise research questions. 
Therefore, we suspect that considering \NP could add value to any interdisciplinary network-science endeavor. 
While this will often result in multi-network analysis, 
even research that sticks to just one perspective would benefit from, 
by default, 
presenting its underlying network space, 
indicating which perspective was chosen, 
and explaining why that perspective was preferred over others. 

\textbf{Limitations.}\quad 
Despite the worked example provided in our case study,
at its core, \NP remains a high-level framework, 
and there is no single, simple way to operationalize the specifics for all domains.
Identifying the relevant dimensions of variation in a given field and mapping them to network-modeling decisions and network-metric parameters will always be a challenging task. 
At the analysis stage,
researchers are confronted with several answers arising from different perspectives, 
and evaluating and relating myriad results in light of these perspectives adds a new level of complexity to the synthesis and interpretation of research findings.

Beyond these conceptual and cognitive challenges,
multi-network analysis increases the computational cost and requires advanced technical expertise to efficiently manage a potentially large number of networks and analytical results (a challenge familiar from multiverse analysis \cite{ganslmeier_estimating_2025}).
Without the technical environment and~tools~that enable \emph{quick and flexible access to the network space}, 
\NP can~consume considerable resources, 
potentially daunting the budding interdisciplinary network~scientist. 

As an inherently interdisciplinary endeavor,
\NP requires in-depth engagement with both domain knowledge and technical expertise 
as well as a frictionless flow of feedback between the two.
This is far from trivial to achieve, 
but it touches the core of interdisciplinary (network) research.
In this sense, \NP provides a skeleton to facilitate interdisciplinary interaction and integration. 

\textbf{Next Steps.}\quad 
Future work could experiment with \NP outside the legal domain.  
This could help the network-science community understand where a pluralist approach might be particularly useful. 
Based on additional insights from diverse domains, 
further studies could also aim to refine \NP as a conceptual framework, 
gradually developing a more detailed, yet generalizable blueprint.

\subsection{Legal Networks}
\textbf{Implications.}\quad
Our analysis showed that legislative units can be described through their references, both individually and in their relation to one another. 
We saw that section profiles based on source and target reference distributions 
can provide a basis for further quantitative research 
and hold the potential to contribute complementary insights to the predominantly normative legal research environment. 
Most importantly, 
adopting \NP in the legal domain not only highlighted where traditional measurement techniques fall short, 
but it also provided a framework for the development of new methods in which the resulting challenges can be overcome. 

\textbf{Limitations.}\quad
The analysis presented in this work is impacted by the narrow scope of the underlying dataset, 
both on the legislative side and on the judicial side.

On the legislative side, all state-level legislation is missing.
This significantly limits our observations because, 
at least in a federal system like Germany, 
most topic-based legislative competencies are typically assigned exclusively to either federal lawmakers or state lawmakers. 
Beyond this coverage gap, 
the legislative data consists of documents in their consolidated form,
i.e., no information about the historical changes made to the documents is available.
This impairs the possibility to include the temporal dimension and inevitably introduces inconsistencies when sections have been changed or moved within the time window under consideration. 

The limitations on the judicial side are of institutional and infrastructural nature.
Decisions taken by courts below the highest federal level are excluded,
i.e., all legal problems contained in this set of court decisions have previously been subject to proceedings within the various prior stages of appeal. 
For a case to make it this far, its questions must pose a certain level of legal complexity and uncertainty, 
and the financial or ideal value of the subject must be high enough to motivate either party to appeal the case to the next instance.
Cases undershooting any of these thresholds, 
such as the simple cases that (presumably) dominate everyday legal practice, 
are not represented in this dataset.

Another constraint stems from the \emph{publication culture} of courts in the German judiciary. 
The data source underlying our case study is provided by the federal government in cooperation with Germany's largest commercial provider of legal data, 
but it contains only ``selected'' decisions from Germany's highest courts. 
Both for this data source and for German courts in general, 
little is known about the procedure underlying the decision to publish a ruling,
and even the publication rate is hardly documented. 
While the highest courts are known to have higher publication rates than other courts, 
empirical research suggests that the overall publication rate (i.e., across all instances) is less than $1\%$ \cite{hamann_blinde_2021}.

\textbf{Next Steps.}\quad
To further advance legal network science, 
future work could consider additional dimensions of variation, thus augmenting the \emph{space of perspectives}.
As the limitations above suggest, the reference relation between jurisprudence and legislation (covered in \Cref{seq:case_studies}) can and should be enriched by including further actors and documents as well as full temporal information.
Regarding the broader goal of generating more complete insights,
subsequent studies should apply \NP to other reference relations (as presented in \Cref{fig:interplay}), 
ultimately combining them to study legal document networks comprehensively without restricting to a specific reference relation. 
Given the limits of one-mode projections, 
a methodological avenue for future work lies in using hypergraphs to more accurately capture references to groups of norms.
Higher-order networks also hold considerable promise for improving the \NP toolkit needed for multi-network analysis more broadly.
Beyond these methodological aspects,
we would welcome the application of \NP to study other jurisdictions or push the frontier of comparative law \cite{coupette_rechtsstruktur_2022}. 

Finally, one overarching bottleneck of legal network science is the limited availability of well-documented, well-preprocessed legal (network) data. 
Removing this bottleneck means building the data infrastructure required to conduct open computational legal research without institutional and geographical boundaries. 
This should be a joined goal of our research community \cite{coupette_sharing_2022}, 
and it would benefit not only legal \emph{network} science but also other fields that rely on legal data---in academia and beyond.

\section*{Data \& Code Availability Statement}

The raw data is provided by the \emph{Federal Ministry of Justice and Consumer Protection}
and can be downloaded and accessed from their free online service; see \Cref{appendix:data} for more information. 
Our code will be maintained on GitHub at \href{https://github.com/telos-lab/network-pluralism}{github.com/telos-lab/network-pluralism}, 
and a reproducibility package will be made available under the following DOI: \href{https://doi.org/10.5281/zenodo.17780510}{10.5281/zenodo.17780510}. 

\section*{Competing Interests}
The authors declare that they have no conflict of interest.

\section*{Author Contributions}
\begin{itemize}[--]
	\item \emph{Titus Pünder}: Conceptualization, data curation, investigation, methodology, software, visualization, writing---original draft.
	\item \emph{Corinna Coupette}: Conceptualization, funding acquisition, methodology, supervision, validation, writing---review and editing.
\end{itemize}

\appendix

\section{Court and Panel Allocation Plans}

In our analysis, we leverage the bipartite relation between actors and their substantive scope of competence as an aid to interpretation and contextualization.
Here, we provide some legal background on this relation and present excerpts from the underlying normative sources. 
\Cref{tab:courts} lists the courts whose decisions are (partly) contained in our dataset. 

It is a constitutional right to know in advance which judge will rule a case (\emph{Right to a Lawful Judge}, Article 101(1) Sentence 2 Basic Law).
The substantive, geographical, and instantiate competence of a \emph{court} is established in statutes.
To bridge the gap between a court as an \emph{institution} and the judge as an \emph{individual},
courts are required to issue an allocation plan that assigns incoming cases to \emph{panels}, 
where each panel is composed of a pre-defined group of judges.
In other words, 
there exist normative rules that assign areas of law to courts and panels.

\begin{table}
    \centering
    \begin{tabular}{llr}
    	\toprule
    \textbf{Court} & \textbf{Area of Law} & \makecell{\textbf{Statutory Basis}} \\ \midrule
    Federal Constitutional Court & Constitutional Law & Article 93(1) Basic Law \\
    Federal Court of Justice & Civil and Criminal Law & Article 95(1) Basic Law \\
    – Civil Panels &–  Civil Law & Case Allocation Plan \\
    – Penal Panels &– Criminal Law & Case Allocation Plan \\
    Federal Administrative Court & Administrative Law & Article 95(1) Basic Law \\
    Federal Fiscal Court & Financial Law & Article 95(1) Basic Law \\
    Federal Labor Court & Labor Law & Article 95(1) Basic Law \\
    Federal Social Court & Social Law & Article 95(1) Basic Law \\
    Federal Patent Court & Patent Law & Article 96(1) Basic Law \\
\bottomrule
  \end{tabular}
  \caption{\textbf{Areas and statutory bases of German Federal Courts.}}\label{tab:courts}
\end{table}

Court allocations plans constitute an extensive corpus of documents.
All documents are publicly available in German. 
To illustrate the assignment of cases to panels and its level of detail and complexity,
in \Cref{fig:court-allocation-plan}, we provide a (translated) excerpt from the allocation plan for the first and second civil panel of the Federal Court of Justice.

\begin{figure}
    \footnotesize
The First Civil Senate is assigned:

\begin{enumerate}
  \item Legal disputes concerning copyright law, publishing law, and design law, including Community design law, as well as general personality rights that are commercially exploited by the entitled person (in the manner of an intellectual property right);

  \item Legal disputes in the field of industrial property protection, insofar as they are not assigned to the Tenth Civil Senate, in particular disputes concerning:
  \begin{enumerate}
    \item trademarks and other distinctive signs (\S 1 of the Trademark Act),
    \item claims under the Act Against Unfair Competition,
    \item claims under the Act on the Protection of Trade Secrets,
    \item name rights, insofar as they concern likelihood of confusion in commercial transactions or disputes concerning domain names.
  \end{enumerate}

 \emph{Items 3--17 omitted.}
\end{enumerate}
The Second Civil Senate is assigned:

\begin{enumerate}

  \item Legal disputes concerning:
  \begin{enumerate}
    \item claims arising from partnership relations (\S\S 705 ff. BGB) and joint ownership communities (\S\S 741 ff. BGB), with the exception of condominium owners' associations, for which the Fifth Civil Senate has jurisdiction,
    \item internal relations of commercial partnerships, silent partnerships, registered cooperatives and associations (including mutual insurance associations), including legal disputes between such entities and their board members or managing directors,
    \item piercing of the corporate veil of members of legal persons (abuse of legal form), unless, with regard to the otherwise applicable law, it appears expedient for the matter to be handled by the Senate competent for that law,
    \item company name law (\S\S 17 ff. HGB), insofar as the First Civil Senate (No. 2(a)) is not competent.\\
\emph{Items (e)--(k)  omitted.}
\end{enumerate}
    \item legal disputes assigned to the Federal Court of Justice under \S 30 sentences 2 and 3 of the Stabilisation Fund Act, unless the Eleventh Civil Senate (No. 6) has jurisdiction,
    \item legal complaints pursuant to \S 70 FamFG in:
    \begin{enumerate}
      \item register matters pursuant to \S 374 FamFG,
      \item corporate law proceedings listed in \S 375 Nos. 1 and 3 to 16 FamFG.
    \end{enumerate}
    \emph{Items 4--7 omitted.}
  \end{enumerate}
\caption{\textbf{Excerpt from the Case Allocation Plan of the Federal Court of Justice.}
The document exhibits a high level of detail. Its internal logic is based on references to statutory units as well as references to other panels.}\label{fig:court-allocation-plan}
\end{figure}

\section{Data and Preprocessing}\label{appendix:data}

\textbf{Legislative Data.}\quad 
The raw legislative data is provided by the \emph{Federal Ministry of Justice and Consumer Protection},
which offers up-to-date downloads of federal statutes and regulations in an XML-format.%
\footnote{A comprehensive table of contents, including all statutes and regulations as well as their download links, can be found at \url{www.gesetze-im-internet.de/gii-toc.xml}.}

The substantive coverage is extensive, including all federal legislation in its consolidated form.
Additions and changes from 1949 up until the last download date (March 16, 2025) are reflected in the XML files.
The raw data files already come in a nested format.
Our preprocessing pipeline simply transforms the original data into sub-items,
capturing the hierarchy and indexing each (sub-)unit up until the lowest available level.
After preprocessing, the legislative data consists of $6,820$ statutes or regulations and $91,758$ sections.

\textbf{Judicial Data.}\quad
	The judicial data is also provided by the \emph{Federal Ministry of Justice and Consumer Protection}.
Every published decision is made available as an XML file.%
\footnote{A comprehensive table of contents, including identifiers of court decisions as well as their download links can be found at \url{www.rechtsprechung-im-internet.de/rii-toc.xml}.}
The dataset contains selected decisions by Germany's highest courts, namely,
the Federal Constitutional Courts,
the Federal Court of Justice (including both civil and penal panels),
the Federal Administrative Court,
the Federal Fiscal Court,
the Federal Labor Court,
the Federal Social Court,
and the Federal Patent Court (see \Cref{tab:courts})---%
where, unfortunately, the selection rule is not made transparent by the data source. 
However, the courts cover a wide range of legal areas, 
such that there is broad substantive coverage.
The time frame ranges from 1st January 2010 up until the last download date (March 16, 2025).

Based on the raw XML files, we extract and index the text of the paragraphs in each decision.
Using the information about the acting panel found in the metadata of each document,
we reconstruct the hierarchy within each court.
From the text,
we extract all references toward any legislative unit indexed in our dataset.
This important preprocessing step builds upon the \emph{QuantLaw} package,
which offers a sophisticated, rule-based reference parser 
tailored toward German legislation \cite{quantlaw}.

\section{Dynamic Counting}\label{appendix:dynamic_counting}

\begin{figure}[h]
\begin{minipage}{.47\textwidth}
    \subfloat[Inconsistent Substructure]{\includegraphics[width=\textwidth]{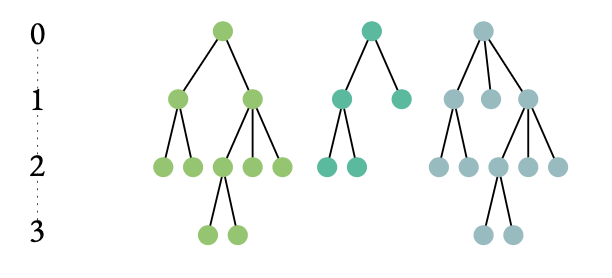}\label{fig:incoherent_depth}}
\end{minipage}
\hfill    
\begin{minipage}{.47\textwidth}
    \subfloat[Inconsistent Reference Targets]{\includegraphics[width=\textwidth]{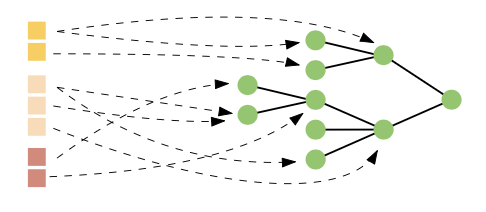}\label{fig:incoherent_references}}
\end{minipage}
    \caption{\textbf{Structural inconsistencies in legal networks.}}\label{fig:incoherence}
\end{figure}

When we treat different levels of granularity across the legislative and judicial side as our dimensions of variation, 
the hierarchical structure itself may be heterogeneous, i.e., inconsistent across levels.
As illustrated in \Cref{fig:incoherence}, this leads to two problems:
inconsistent substructure and inconsistent reference targets. 

When left untreated, 
these problems distort the networks derived from our base network.
Without further adaption,
setting levels of granularity
would not only change the dimensions of the network (a desired effect) 
but also the number of references captured (an undesired effect),
as references to units above or below the specific granularity level would be omitted.
To address these problems, 
we suggest separating the reference counting from generating the network model. 
That is, we no longer construct networks directly from the raw data
but rather fix inconsistencies before network construction, 
allowing us to generate specific networks without information loss.

\begin{figure}[h]
\centering
\includegraphics[ width=0.8\textwidth]{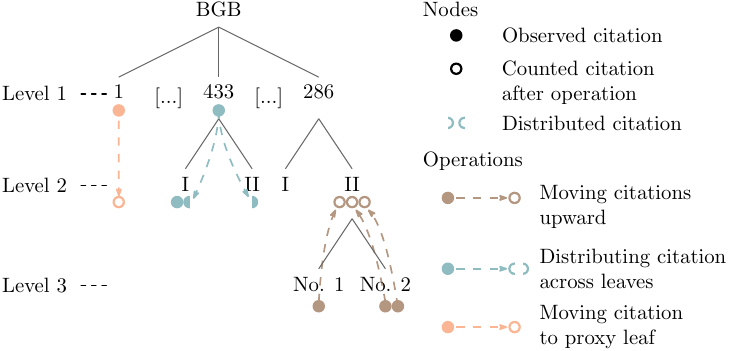}
\vspace{1em}
\caption{\textbf{Dynamic counting of references.} We show the primitive operations underlying the algorithm we use to resolve inconsistencies across reference levels.
}\label{fig:dynamic_counting_operations}
\end{figure}

To achieve this,
we provide a \emph{Dynamic Counting} algorithm 
that separately deals with both types of inconsistency.
Its goal is to distribute all references downward to the overall lowest level.
This is achieved by creating proxy nodes to ensure all (sub)sections exists up until the lowest level,
and by splitting references at higher levels to evenly distribute them at lower levels.
Thus, information from references to higher levels is proportionally propagated to lower (= more granular) levels. 
This process is then reversed when the network is considered at a higher level, 
where lower-level references are aggregated up the hierarchy until the desired resolution level is reached.  
\Cref{fig:dynamic_counting_operations} illustrates the process. 

From a computational perspective, 
our algorithm combines two classic operations on trees that are often used in distributed computing: 
a downward broadcast operation that distributes references to the most granular level to enable lossless network generation at all levels of resolution, 
and an upward aggregation operation that, on demand, collects references from subtrees at the desired level of granularity.

\section{Statute Abbreviations}\label{appendix:abbreviations}

The following table resolves and translates all statute abbreviations used in our figures or tables. 
\begin{table}[h!]
    \centering
    \begin{tabular}{l@{\hspace*{0.5em}}l@{\hspace*{0.5em}}l@{\hspace*{0.5em}}l}
    	\toprule
        \textbf{Abbr.} & \textbf{Original German Title} & \textbf{Translation} & \textbf{Figures} \\ \midrule
        BGB     & Bürgerliches Gesetzbuch      & Civil Code                               & \ref{fig:in_degree_and_distributions}, \ref{fig:bar_combarison_granularity_433}, \ref{fig:bar_comparison_granularity}, \ref{fig:bar_combarison_granularity_187} \\ 
        
        FGO     & Finanzgerichtsordnung        & Code of Fiscal Court Procedure           &  \ref{fig:in_degree_and_distributions}, Table \ref{table:overrepresentation} \\ 
        
        GG      & Grundgesetz                  & Basic Law                                 &  \ref{fig:in_degree_and_distributions}, \ref{fig:bar_combarison_granularity_187}, Table \ref{table:overrepresentation}\\ 
        GKG     & Gerichtskostengesetz         & Court Fees Act                            &  Table \ref{table:overrepresentation} \\ 
        MarkenG & Markengesetz                 & Trade Mark Act                            & \ref{fig:in_degree_and_distributions} \\ 
        StGB    & Strafgesetzbuch              & Criminal Code                             &  \ref{fig:bar_combarison_granularity_187}\\ 
        StPO    & Strafprozessordnung          & Code of Criminal Procedure                &  \ref{fig:in_degree_and_distributions}, \ref{fig:bar_combarison_granularity_187} \\ 
        VwGO    & Verwaltungsgerichtsordnung   & Code of Administrative Court Procedure    & \ref{fig:in_degree_and_distributions}, \ref{fig:bar_combarison_granularity_187}, Table \ref{table:overrepresentation} \\ 
        WBO     & Wehrbeschwerdeordnung        & Military Complaints Code                  & \ref{fig:bar_combarison_granularity_187} \\ 
        ZPO     & Zivilprozessordnung          & Code of Civil Procedure & \ref{fig:in_degree_and_distributions}, \ref{fig:bar_combarison_granularity_433}, \ref{fig:bar_combarison_granularity_187}, Table \ref{table:overrepresentation} \\
        \bottomrule
    \end{tabular}
\end{table}

\bibliography{tools,references}

\end{document}